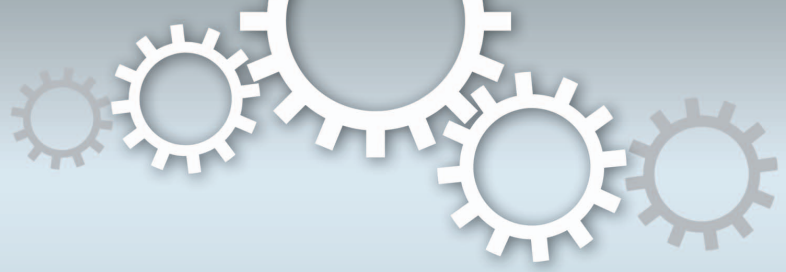

**OPEN**

# Analysis of low-field isotropic vortex glass containing vortex groups in YBa$_2$Cu$_3$O$_{7-x}$ thin films visualized by scanning SQUID microscopy

Frederick S. Wells[1], Alexey V. Pan[1], X. Renshaw Wang[2]*, Sergey A. Fedoseev[1] & Hans Hilgenkamp[2]

[1]Institute for Superconducting and Electronic Materials, University of Wollongong, Northfields Avenue, Wollongong, NSW 2500, Australia, [2]Faculty of Science and Technology and MESA + Institute for Nanotechnology, University of Twente, P.O. Box 217, 7500 AE Enschede, The Netherlands.

Correspondence and
requests for materials
should be addressed to
AVP. (pan@uow.edu.
au)

* Current address:
Electrochemical
Energy Laboratory,
Massachusetts Institute
of Technology,
Cambridge, MA
02139, USA

The glass-like vortex distribution in pulsed laser deposited YBa$_2$Cu$_3$O$_{7-x}$ thin films is observed by scanning superconducting quantum interference device microscopy and analysed for ordering after cooling in magnetic fields significantly smaller than the Earth's field. Autocorrelation calculations on this distribution show a weak short-range positional order, while Delaunay triangulation shows a near-complete lack of orientational order. The distribution of these vortices is finally characterised as an isotropic vortex glass. Abnormally closely spaced groups of vortices, which are statistically unlikely to occur, are observed above a threshold magnetic field. The origin of these groups is discussed, but will require further investigation.

S canning SQUID Microscopy (SSM) is one of the most powerful techniques for measuring the local magnetic field above the surface of a sample. SQUIDs (Superconducting Quantum Interference Devices) are the most sensitive magnetometers in existence, and the resolution obtained by this technique has been shown to be greater than that obtained by many other local magnetic field measurement techniques[1].

Autocorrelation and Delaunay triangulation techniques have been applied to the scanning SQUID images of vortices in a-MoGe superconducting thin films with relatively weak pinning, to quantify deviations from the triangular vortex lattice around pinning sites[2]. In YBa$_2$Cu$_3$O$_{7-x}$ (YBCO) superconducting thin films, scanning SQUID images have been acquired for only a few individual vortices[3] for current mapping. Larger-scale vortex distributions across YBCO films had not heretofore been examined. YBCO thin films grown by pulsed laser deposition (PLD) have strong pinning dominated by the out-of-plane dislocations[4–6], enabling a vortex glass state. The distribution of vortices in the glass state differs strongly from the vortex lattice[7,8]. The glass state is a stable vortex arrangement characterised by quenched disorder, which arises due to pinning.

Vortex glass states can be classified according to experimental conditions (field strength, temperature, magnetic history, etc.), the type of material (anisotropy, dimensionality, etc.), and type(s) of available effective pinning centres[8–12]. Glass states may even transform from one type to another, depending on these parameters. For example, weak pinning may be realised as a small number of weak pinning sites or by enabling strong thermal activations in the proximity of the superconducting transition temperature, which may lead to the so-called positional vortex glass. The typical example would be the 3D Bragg-like glass phase, exhibiting no short-range order, but still preserving a long-range ordering. In the case of PLD YBCO thin films, possessing strong out-of-plane columnar defects (dislocations), isotropic Bose-like glass[10,11] having no ordering would be expected, but was never observed. Vortex glasses can often be hexatic in nature, showing sixfold orientational order[9]. Both isotropic and hexatic glassy states have been thoroughly examined by simulations[13,14], but transitions between these states have not been directly observed.

In this work, scanning SQUID microscopy has been employed to visualise and then to characterise field-cooled vortex states in PLD-deposited YBCO thin films in low fields ($\mu$T range). The ordering of the vortices has been analysed through autocorrelation and Delaunay triangulation techniques. We show that (i) although the most vortices have six nearest neighbours, their angular distribution does not show the orientational order of a hexatic vortex glass; (ii) although we can classify our observations as an isotropic vortex glass, we have also found a signature of short-range ordering. In addition, we have unexpectedly found groups of closely spaced vortices which show an appearance of interacting currents. These groups are also examined and discussed.

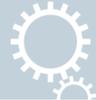

## Methods

A scanning SQUID microscope consisting of a $\mu$-SQUID magnetometer with a 3 $\mu$m × 5 $\mu$m pickup loop is raster scanned at a distance of approximately 5 $\mu$m from the YBCO sample surface at an angle of 30°. The SSM measurements were performed in a crystat equipped with a $\mu$-metal shield with an approximate shielding factor of 25. This reduces the Earth's magnetic field to a constant background field during cooling and scanning. The vertical component of this background field was measured to be approximately 2 $\mu$T by a Bartington Mag-03MS three-axis magnetic field sensor, with other components negligibly small, and this background field value was confirmed as 2.73 $\mu$T by preliminary scanning SQUID measurements on our samples. That is the field of −2.73 $\mu$T was found to reduce the number of observed vortices to a minimum, this was also found to be the field value at which the vortex direction reversed. The root-mean-square (RMS) variation of the background field was found to be less than 30 nT by antiferromagnetic scanning SQUID microscopy. All subsequent field values stated in this paper are given after compensating for this background field.

All measurements were taken at a temperature of 4.2 K in the field-cooled state, with applied fields in the range 0.1 $\mu T < B_a < 5.5$ $\mu T$ perpendicular to the film's surface.

Local current distribution in the samples was calculated from the magnetic field data using a program[15] based on an inverse Biot-Savart procedure[16,17]. The arrangement of vortices was further analysed by autocorrelation and Delaunay triangulation based on vortex positions.

The YBa$_2$Cu$_3$O$_{7-x}$ thin films used in this work have been grown by pulsed laser deposition[18,19] with the thickness of ~200 nm. The critical temperature ($T_c$) of the films has been measured by magnetisation measurements to be 90.0 ± 0.5 K. The surface of the films has been observed by atomic force microscopy, showing an average grain size of about 200 nm.

## Scanning squid microscopy

Figure 1 shows the local magnetic field data obtained by the scanning SQUID microscope. The brightness of each point in the image shows the magnetic field strength at the corresponding point above the sample. Vortices are seen as round dark spots over the right-hand side of the images. An identifiable position at the edge of the film was chosen for scanning to ensure that repeat scans were taken at the same position on the film. This edge is seen at the left side of the images.

Since the SQUID magnetometer scans at a constant height of 5 $\mu$m above the sample, the magnetic features observed are those of the stray field. In this paper the term "stray field" refers to the observed magnetic field at the scan height as opposed to the field directly at the film's surface, and "stray current" refers to the current in the film as calculated from the stray field. This distance from the sample surface increases the apparent size of the vortices in Fig. 1. The vortices also appear slightly asymmetrical in Fig. 1 due to the tilt of the SQUID pick-up loop with respect to the field direction.

Figure 2 shows the current distribution in the sample calculated from the magnetic field data of figure 1. The brightness of each point in the image is proportional to the magnitude of current at the corresponding point in the sample. The dark spots seen throughout the sample and the bright regions around them are the current-free vortex cores and the circulating current of the vortices, respectively.

The distance between the midpoints of neighbouring vortices has been determined from the field maps in Fig. 1. At $B_a \simeq 6.93$ $\mu$T, the average intervortex spacing is 32 $\mu$m, with a significant spread in nearest neighbour distances as expected in glassy distributions. However, there were a disproportionately large number of vortices with nearest neighbour distances in the range of ~ 15 $\mu$m. The groups of these closely spaced vortices in Fig. 2 are mapped to have overlapping *stray* supercurrents that are continuous around the perimeter of the whole group. However, in this strongly diluted vortex regime the magnetic field penetration depth ($\lambda$)[20] and the individual vortex depinning radius in YBCO films are of the order of 0.5 $\mu$m[21,22], being too small to have any profound effect at such large intervortex distance within the group[21,22]. Thus, the supercurrent overlap is probably due to the spread of stray fields at the SQUID scanning height[23].

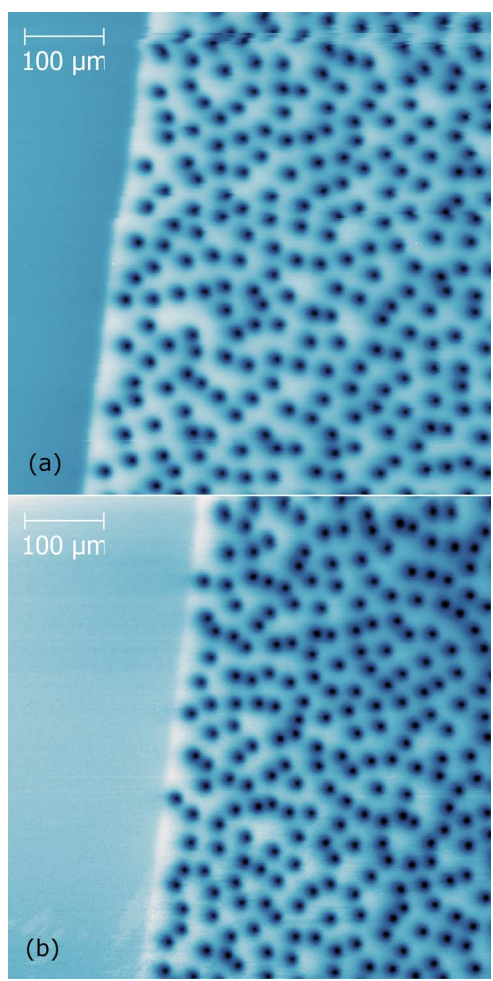

**Figure 1 | Images of vortices in 200 nm thick YBCO film taken by Scanning SQUID Microscopy after field cooling at 6.93 $\mu$T to 4 K. (b) is taken after heating above $T_c$ and re-cooling. The sample edge at the left side of the images is used as a reference for scan location.**

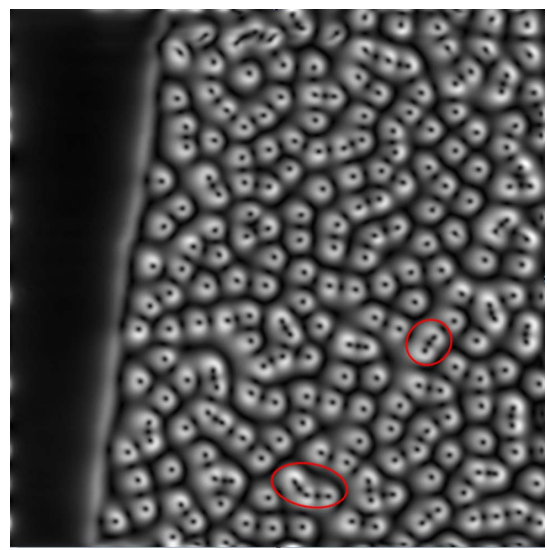

**Figure 2 | Supercurrents calculated from the field map in Fig. 1(a). Some closely-spaced vortex groups are highlighted by the circles.**

 

However, the spread of stray fields does not explain vortex grouping. The onset of such grouping behaviour is observed at $B_a \gtrsim 2$ $\mu$T. At lower fields, the vortices are spaced very far apart and no connectedness of circulating supercurrents was mapped. Multiple images were taken for each magnetic field value after heating the sample above its critical temperature and re-cooling to reset the vortex distribution. If the vortex grouping occurred due to particular strong pinning centres, a reproducibility of the locations of the groups would be expected upon re-cooling. In general, this did not occur. The positions of the groups appear randomly and independently distributed for each cooling.

Hence, it can be inferred that the vortex groups are not formed around any particular strong pinning centres in the films. It is quite possible that it occurs due to the much diluted (gas-like) vortex regime at very low fields. Being field-cooled through the transition temperature, magnetic flux quantisation occurs in this gas-like state with negligible intervortex interactions. Indeed, upon progressive cooling the thermal activations statistically drive vortices into the nearest pinning positions, which once frozen at lower temperatures exhibit the observed configurations. However, it is still statistically unlikely that such grouping should occur with one pronounced intervortex spacing ($\sim$15–20 $\mu$m).

The prevalence of a particular vortex spacing in the groups may be an indicator of a straightforward mechanism such as film inhomogenity or strong demagnetisation factor leading to regions of heightened field. However, the position of the groups is not reproducible upon re-cooling and few vortices are grouped together. If there were only certain larger inhomogenities in the films, the group positions would be reproducible. And if the inhomogeneities were evenly distributed throughout the film, most of the vortices would be combined in groups. Hence, we can largely rule out inhomogeneity in the film as the cause of vortex grouping. As the films were cooled in field, the demagnetising factor would be negligible at the field of a few $\mu$T, hence it is also unlikely that this is the reason for the observed grouping. Thus, further study is required to identify the group formation mechanism.

The size of vortices was determined by plotting field profiles across vortices (data points shown in figure 3). A Gaussian profile (the solid curve in Fig. 3) was fitted to the stray field of vortices, and the average full width half maximum (FWHM) was found to be 6 $\mu$m. This is significantly larger than the expected size of $2\lambda \sim 0.5$ $\mu$m at the film surface. The large size of vortex field profiles shows a significant spread of stray field above the film. However, the size may also be influenced by the spatial resolution of the scanning SQUID apparatus, which is limited by the size of the pickup loop. In any case, the

$2\lambda$ size of vortices is about $>2$ to 6 times smaller compared to the intervortex spacing, indicating negligible intervortex interactions in our frozen diluted vortex glass.

Interesting supercurrent profiles are obtained for the shielding currents around vortices. The expected current-free region is observed at the centre of each vortex, although the average FWHM of the *stray* current-free region is seen to be 3.2 $\mu$m. This is far too large compared to the expected vortex core diameter of $2\xi$, which is about 3 to 4 nm[24], where $\xi$ is the coherence length. Outside this region, the stray shielding currents peak sharply before gradually fading to zero at some distance from the vortex core. The most interesting current mapping result is for the groups of vortices: Negligible current is observed between the grouped vortices (Fig. 2), suggesting a connectedness of supercurrents around the whole group. This may simply be the result of the superposition of the stray fields from the oppositely directed shielding supercurrents around neighbouring vortices in the groups.

However, a more precise analysis is impossible due to the strongly exaggerated dimensions observed due to stray field spread above the sample. The field directly at the sample surface cannot be calculated from the stray field due to the difficulty in precisely determining the distance from sensor to sample in the scanning SQUID apparatus. If this calculation could be carried out with a level of precision that surpasses the current equipment limitations, then any connectedness of the supercurrent profiles in the vortex groups could be examined more rigorously, and this appearance of connectedness may disappear.

## Analysis of vortex arrangements

To characterise the vortex phase, an autocorrelation process is employed. A spatial autocorrelation $AC(\vec{r})$ for some displacement $\vec{r}$ maps the average correlation of every point in a set of data to the corresponding point separated by this displacement.

$$AC(\vec{r}) = \int B(\vec{r})B(\vec{r} + \vec{r}')d\vec{r}' \qquad (1)$$

The integral in Eq. 1 is over the image area. When applied to magnetic field maps, the autocorrelation value for some displacement $\vec{r}$ gives the probability that two points separated by $\vec{r}$ have the same field value. Therefore, the brightness gives the probability of a vortex occurring with separation $\vec{r}$ from another vortex.

Figure 4(b) shows the autocorrelation for a modelled perfect triangular lattice composed of a finite number of points of finite size. This represents the "ideal" defect-free vortex lattice in a superconducting film. Figure 4(a) shows the calculated autocorrelation from a field map of the vortex distributions observed by scanning SQUID shown in Fig. 1(b). Compared to the "ideal" vortex lattice, the glassiness of the vortex distribution in our films is evident due to the lack of regularity in this autocorrelation. However, the somewhat brighter ring around the central point may indicate retention of some weak short-range order. Further from the centre, no ordering can be seen, which implies a lack of long-range order, confirming the glassy nature of this diluted vortex distribution.

To further classify this glass as hexatic or isotropic, the orientational order must be analysed, and this is achieved by Delaunay triangulation. Delaunay triangulation is a procedure which maps connections between points to form triangles such that no point lies within the circumcircle of any triangle. This method effectively maps connections between nearest neighbours.

The visualisation of Delaunay triangulation applied to the field map in Fig. 1(a) is shown in Fig. 5(a). From this analysis, the distribution of the number of nearest neighbours to each vortex is obtained and plotted in Fig. 5(b) for three independent scans under identical field conditions. The number of nearest neighbours to each vortex is equivalent to the number of triangles connecting to it. This consideration is not valid for the vortices at the edges of the analysed area. Thus, all vortices with a mismatching number of neighbours and triangles, as well as their immediate neighbours have been excluded from the calculations to prevent edge effects.

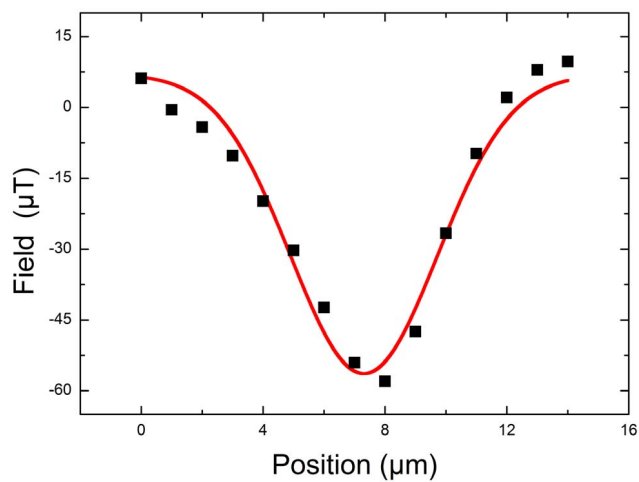

**Figure 3 | The field profile across a single vortex along with its Gaussian fit (solid line).**

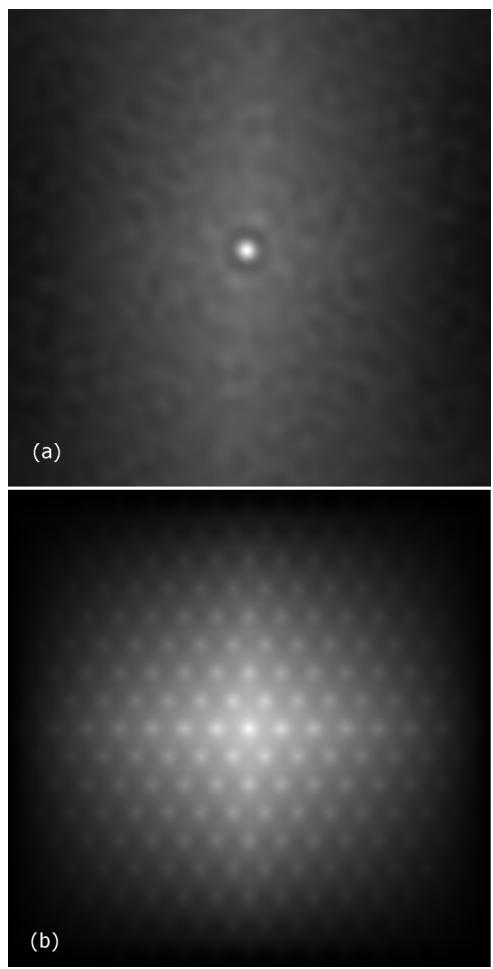

**Figure 4** | (a) The visualisation of the autocorrelation function from Fig. 1(a). The brightness of a point at some position $\vec{r}$ from the centre represents the autocorrelation for the corresponding displacement. (b) Modelled autocorrelation for a finite "ideal" triangular lattice.

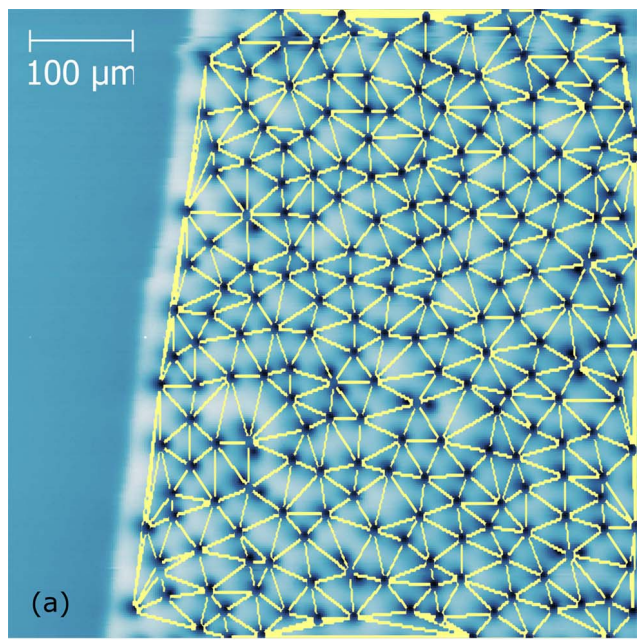

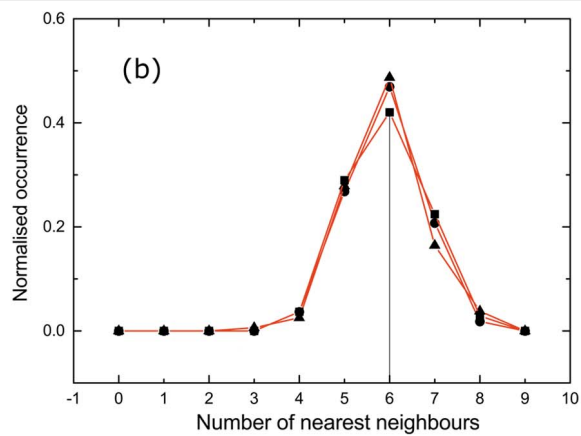

**Figure 5** | (a) Delaunay triangulation mapped onto the vortex distribution shown in Fig. 1(a). (b) The normalised plot of the number of nearest neighbours to each vortex, calculated after three re-warming and re-cooling at the same conditions.

The triangulation has been calculated based on positions of vortices found by applying particle detection to the images; the coordinates of the centre of each particle are taken as the vortex position. This calculation method is generally accurate, but fails when pairs of vortices are so closely spaced that the program recognizes them as a single particle. This situation can be seen in several places in Fig. 5(a), where there is more than one vortex at a single node of the triangulation. It occurs approximately once for every 40 vortices and tends to underestimate the number of nearest neighbours for some vortex pairs, and overestimate for others. This may cause a slight spread in the results of Fig. 5(b). Hence, the true distribution of the nearest neighbours is expected to peak more sharply at 6.

Since most vortices are seen to have six nearest neighbours, orientational order is investigated using the hexatic order parameter of the lattice $|\psi_6|^2$, given by

$$\psi_6 = \frac{1}{N_{int}} \sum_i \frac{1}{n(i)} \sum_j e^{6i\theta_{ij}}, \qquad (2)$$

where $N_{int}$ is the number of vortices (excluding the vortices on the edges), $n(i)$ is the number of nearest neighbours to a vortex $i$ and $\theta_{ij}$ is the bond angle between neighbouring vortices $i$ and $j$ relative to some arbitrary axis.

From Eq. 2, it can be readily seen that a perfect triangular lattice has a hexagonal order parameter $\psi_6 = 1$ due to its sixfold rotational symmetry. The vortex distribution examined in Fig. 5(a) has been found to have a hexagonal order parameter $\psi_6$ between 0.01 and 0.001, implying a near-complete lack of orientational order.

## Summary

In summary, the observed distribution of vortices in PLD YBCO thin films in this extremely diluted field-cooled vortex regime in low fields of the $\mu$T-range by scanning SQUID microscopy can therefore be characterised as an isotropic vortex glass. The autocorrelation technique has demonstrated the lack of long-range order as compared to a perfect lattice, while a fading signature of a weak short-range order has been retained. Employing the Delaunay triangulation technique, the vortex glass has then been shown to be isotropic although retaining six nearest neighbours. The subsequent calculation of the hexatic orientational order parameter has been found to be very low.

None of the vortex glass features found in the films is at odds with possible expectations and existing theories, but we cannot favour any specific further classification of the glassy state. Although, it is tempt-ing to assign Bose-like features to this glass state due to the presence of strong columnar-like defects (out-of-plane dislocations) in the films. This indeterminacy is most likely due to the highly diluted vortex regime formed by cooling in very low fields. Being field-cooled through the transition temperature, magnetic flux quantisation occurs in a rather gas-like state with negligible intervortex interactions, so that upon progressive cooling the thermal activations statistically drive vortices into the nearest strong pinning positions, which once frozen exhibit the observed configurations at lower temperatures.

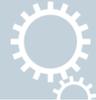

Within this isotropic vortex glass, some unexpectedly closely spaced groups of vortices are observed with pronounced intervortex distances of around 15 $\mu$m at $B_a = 6.93$ $\mu$T, which is around half the average vortex spacing. This grouping is highly unlikely under statistical thermally driven vortex redistribution upon cooling and the groups did not appear to be associated with any particularly strong-pinning defects (or inhomogeneities) in the sample. Thus, the reason for this type of grouping is ultimately unclear, and may be the subject of further work.

1. Kirtley, J. R. & Wikswo Jr., J. P. Scanning SQUID microscopy. *Annual Reviews of Materials Science.* **29**, 117 (1999).
2. Plourde, B. L. T. *et al.* Vortex distributions near surface steps observed by scanning SQUID microscopy. *Phys. Rev. B.* **66**, 054529 (2002).
3. Sugimoto, A., Yamaguchi, T. & Iguchi, I. Supercurrent distribution in high-$T_c$ superconducting $Y Ba_2Cu_3O_{7-y}$ thin films by scanning superconducting quantum interference device microscopy. *Appl. Phys. Lett.* **77**, 3069 (2000).
4. Pan, V. *et al.* Supercurrent transport in $YBa_2Cu_3O_7$ epitaxial thin films in a dc magnetic field. *Phys. Rev. B* **73**, 054508 (2006).
5. Pan, A. V. *et al.* Quantitative Description of Critical Current Density in YBCO Films and Multilayers. *IEEE Trans. Supercond.* **19**, 3391 (2009).
6. Golovchanskiy, I. *et al.* Rectifying differences in transport, dynamic, and quasi-equilibrium measurements of critical current density. *J. Appl. Phys.* **114**, 163910 (2013).
7. Fisher, D. S., Fisher, M. P. A. & Huse, D. Thermal fluctuations, quenched disorder, phase transitions, and transport in type-II superconductors. *Phys. Rev. B.* **43**, 130 (1991).
8. Blatter, G., Feigel'man, M. V., Geshkenbein, V. B., Larkin, A. I. & Vinkour, V. M. Vortices in high temperature superconductors. *Rev. Mod. Phys.* **66**, 054529 (2002).
9. Natterman, T. & Scheidl, S. Vortex-glass phases in type-II superconductors. *Adv. Phys.* **49**, 607 (2000).
10. Nelson, D. R. & Vinokur, V. M. Boson localization and correlated pinning of superconducting vortex arrays. *Physical Review B.* **48**, 13060 (1993).
11. Lindmar, J. & Wallin, M. Critical properties of Bose-glass superconductors. *Europhys. Lett.* **47**, 494 (1999).
12. Nonomura, Y. & Hu, X. Possible Bragg-Bose glass phase in vortex states of high-$T_c$ superconductors with sparse and weak columnar defects. *Europhys. Lett.* **65**, 533 (2004).
13. Ryu, S. & Stroud, D. First-order melting and dynamics of flux lines in a model for $YBa_2Cu_3O_{7-\delta}$. *Phys. Rev. B.* **54**, 1320–1333 (1996).
14. Bishop, D. J. *et al.* Observation of an hexatic vortex glass in flux lattices of the high-$T_c$ superconductor $Bi_{2.1}Sr_{1.9}Ca_{0.9}Cu_2O_{8+\delta}$. *Physica B.* **169**, 72–79 (1991).
15. Wells, F. S. Magneto-optical imaging and current profiling on superconductors, Honours Thesis, University of Wollongong (2011).
16. Roth, B. J., Sepulveda, N. G. & Wiskwo, J. P. Using a magnetometer to image a two-dimensional current distribution. *J. Appl. Phys.* **65**, 361 (1988).
17. Jooss, C., Albrech, J., Kuhn, H., Leonhardt, S. & Kronmüller, H. Magneto-optical studies of current distributions in high-$T_c$ superconductors. *Rep. Prog. Phys.* **65**, 651 (2002).
18. Pan, A. V., Pysarenko, S. & Dou, S. X. Drastic improvement of surface structure and current-carrying ability in $YBa_2Cu_3O_7$ films by introducing multilayered structure. *Appl. Phys. Lett.* **88**, 232506 (2006).
19. Pan, A. V., Pysarenko, S. V., Wexler, D., Rubanov, S. & Dou, S. X. Multilayering and Ag-Doping for Properties and Performance Enhancement in $YBa_2Cu_3O_7$ Films. *IEEE Trans. Appl. Supercond.* **17**, 3585 (2007).
20. de Vaulchie, L. A. *et al.* Linear temperature variation of the penetration depth in $YBa_2Cu_3O_{7-\delta}$ thin films. *Europhys. Lett.* **33**, 153 (1996).
21. Pan, A. V. *et al.* Thermally activated depinning of individual vortices in $YBa_2Cu_3O_7$ superconducting films. *Physica C.* **407**, 1016 (2004).
22. Pan, A. V. & Dou, S. X. Comparison of small-field behavior in $MgB_2$, Low- and high-temperature superconductors. *Phys. Rev. B.* **73**, 052506 (2006).
23. Tsuchiya, Y., Nakajima, Y. & Tamegai, T. Development of surface magneto-optical imaging method. *Physica C.* **470**, 1123–1125 (2010).
24. Pan, V. M. & Pan, A. V. Vortex matter in superconductors. *Fiz. Niz. Temp.* (Rus.) **27**, 991 (2001) *Low Temp. Phys.* **27**, 732 (2001).

## Acknowledgments

This work is supported by the Australian Research Council within Discovery Project (DP110100398). X.R.W. thanks the support from the Dutch NWO foundation through a Rubicon grant (2011, 680-50-1114). The work has benefited from fruitful discussions with I.A. Golovchanskiy.

## Author contributions

F.S.W. acquired the scanning SQUID data, performed all calculations, wrote the manuscript, and prepared the figures. A.V.P. initiated and drove the work, provided YBCO films, wrote and revised the manuscript. X.R.W. acquired the scanning SQUID data. S.A.F. provided YBCO films. H.H. facilitated the scanning SQUID microscope observations, discussed the data obtained, contributed to the data analysis, and revised the manuscript.

## Additional information
**Competing financial interests:** The authors declare no competing financial interests.

**How to cite this article:** Wells, F.S., Pan, A.V., Wang, X.R., Fedoseev, S.A. & Hilgenkamp, H. Analysis of low-field isotropic vortex glass containing vortex groups in $YBa_2Cu_3O_{7-x}$ thin films visualized by scanning SQUID microscopy. *Sci. Rep.* **5**, 8677; DOI:10.1038/srep08677 (2015).